\documentclass[%
reprint,
superscriptaddress,
amsmath,amssymb,
aps
]{revtex4-2}

\usepackage{graphicx}
\usepackage{dcolumn}
\usepackage{bm}
\usepackage{hyperref}
\usepackage{layouts}
\usepackage{xcolor}
\usepackage{comment}
\usepackage{multirow} 
\usepackage{gensymb}
\usepackage{braket}

\begin{document}

\title{Investigation of the Stark Effect on a Centrosymmetric Quantum Emitter in Diamond}

\author{Lorenzo De Santis}
\affiliation{Department of Electrical Engineering and Computer Science, Massachusetts Institute of Technology, Cambridge, Massachusetts 02139, USA}

\author{Matthew E. Trusheim}
\affiliation{Department of Electrical Engineering and Computer Science, Massachusetts Institute of Technology, Cambridge, Massachusetts 02139, USA}

\author{Kevin C. Chen}
\affiliation{Department of Electrical Engineering and Computer Science, Massachusetts Institute of Technology, Cambridge, Massachusetts 02139, USA}

\author{Dirk R. Englund}
\affiliation{Department of Electrical Engineering and Computer Science, Massachusetts Institute of Technology, Cambridge, Massachusetts 02139, USA}

\begin{abstract}
Quantum emitters in diamond are leading optically-accessible solid-state qubits. Among these, Group IV-vacancy defect centers have attracted great interest as coherent and stable optical interfaces to long-lived spin states. Theory indicates that their inversion symmetry provides first-order insensitivity to stray electric fields, a common limitation for optical coherence in any host material. Here we experimentally quantify this electric field dependence via an external electric field applied to individual tin-vacancy (SnV) centers in diamond. These measurements reveal that the permanent electric dipole moment and polarizability are at least four orders of magnitude smaller than for the diamond nitrogen vacancy (NV) centers, representing the first direct measurement of the inversion symmetry protection of a Group IV defect in diamond. Moreover, we show that by modulating the electric-field-induced dipole we can use the SnV as a nanoscale probe of local electric field noise, and we employ this technique to highlight the effect of spectral diffusion on the SnV.
\end{abstract}

\maketitle

Quantum emitters in diamond have emerged as leading solid-state quantum memories. The nitrogen vacancy (NV) center, in particular, has been used for basic quantum network demonstrations \cite{Hensen2015,Childress2006} including on-demand entanglement \cite{Humphreys2018}. Despite the NV's excellent spin properties, its optical interface is inefficient: due to strong electron-phonon interactions, only a small fraction of emission occurs into the spin-correlated zero-phonon line (low Debye-Waller factor) \cite{Gruber1997}. Integration into photonic nanostructures can improve this branching ratio, but at the cost of linewidth broadening and spectral diffusion because of the NV's sensitivity to electric field fluctuations, which are particularly strong near material interfaces \cite{Faraon2012,Li2015}.
These challenges have sparked strong interest towards inversion-symmetric Group IV-Vacancy quantum emitters (SiV, GeV, SnV and PbV), whose Debye-Waller factor can be an order of magnitude greater than in the NV. Additionally, these emitters have been shown to have low spectral diffusion and lifetime-limited emission even in nanostructures \cite{Evans2016a}.
Such robustness is attributed to the split vacancy configuration of the emitters with D$_{3d}$ symmetry, which has been predicted to produce a first order insensitivity to electric fields \cite{Thiering2018a}.
This property enabled the demonstration of high QED cooperativities for cavity-coupled emitters \cite{Evans2018a}, wide spectral tuning of their emission energies \cite{MacHielse2019}, and large-scale integration of defects in waveguide arrays \cite{Wan2019}.
Despite being one of its key advantages, no measurement of the first-order electric field insensitivity of a Group IV emitter has been reported to date.
In this work, we directly test the expected insensitivity to electric fields by investigating the Stark effect on a single negatively charged SnV emitter.
We confirm the absence of a significant permanent electric dipole moment, resulting from the inversion symmetry of the defect, as well as an extremely low polarizability --- orders of magnitude lower than for the NV center.
Finally, we use the electric-field-induced dipole of the SnV defect as a nanoscale probe to estimate the electric field noise in its vicinity.

\begin{figure}[b!]
	\includegraphics[width=0.5\textwidth]{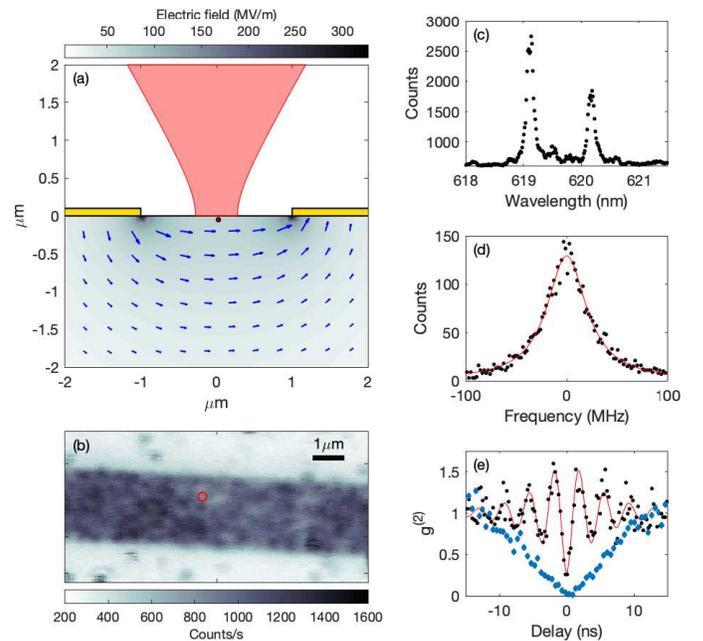}
	\caption{\label{fig:sample_geometry} (a) Side view of the sample layout, showing two gold electrodes and the expected external electric field distribution as simulated with COMSOL for an external bias of 200 V. The emitter location (expected 76 nm below the suface) is indicated by a red circle. (b) Confocal photoluminescence scan of the sample surface, showing an ensemble of defect centers in the exposed diamond, enclosed between two electrodes. (c) Typical emission spectrum obtained under non-resonant excitation. (d) PLE spectrum of the transition to the lower ground state of an SnV, showing a linewidth of 50MHz (e) second order correlation measurement on a single defect, performed at high power (black dots) and low power (blue diamonds).}
\end{figure}

We consider the negatively-charged SnV in diamond as a prototypical Group-IV emitter. They share qualitatively identical electronic structure consisting of four electronic orbitals connected by four optical transitions \cite{Thiering2018a}. The $D_{3d}$ symmetry that gives rise to the predicted first-order insensitivity is preserved \cite{Wahl2020,Gorlitz2020,Rugar2020}. Moreover, the large ground-state orbital splitting of the SnV (850 GHz for SnV compared to 50 GHz for SiV) gives it the potential for long spin coherence above dilution refrigerator temperatures, a key consideration for use as a quantum memory \cite{Iwasaki2017,Trusheim2020}.

The SnV emitters analyzed here are created in a CVD-grown type IIa diamond through ion implantation and subsequent vacuum annealing. To probe the electric field response, we deposit on the surface an interdigitated electrode structure, with an inter-electrode spacing ranging from $2~\mu m$ to $10~\mu m$ (see supplementary material for details of the sample fabrication).
This allows us to apply an electric field aligned with the cryostalline [110] axis as depicted in Figure \ref{fig:sample_geometry}(a).
We characterized the SnV emitters using a custom-built confocal setup, where the sample is kept at a temperature of 4 K inside a closed-cycle liquid helium cryostat (Montana Instruments), while the electric field applied to the emitters is controlled by an external DC voltage source, supplying up to $\pm 200$~V to the electrodes with a negligible leakage current (details in Supplementary).
A spatial photoluminescence (PL) emission map obtained with a 515 nm pump laser is shown in Figure \ref{fig:sample_geometry}(b). The emitted PL spectrum (Figure \ref{fig:sample_geometry}(c)) consists of many spectral lines, indicating an ensemble of emitters within the optical diffraction limit.
To investigate single SnV defects, we use photoluminescence excitation (PLE), by scanning the frequency of a narrow-band laser across the SnV transition while detecting the emitted photons in the red-detuned phonon sideband.
Figure \ref{fig:sample_geometry}(d) plots the result of a PLE measurement, from which we can extract a single-transition linewidth of $47\pm2$ MHz.
By fixing the excitation laser on resonance with the transition, we perform a second-order correlation measurement as shown in Figure \ref{fig:sample_geometry}(e). At low excitation power we obtain a value of $g^{(2)}(0)=0.03$, confirming the single-photon nature of the collected PL, as well as an optical $T_1$ lifetime of $6.0\pm0.5$ ns corresponding to a lifetime-limited linewidth of $27\pm2$ MHz. Under strong driving, the $g^{(2)}$ measurement reaveals Rabi oscillations with a phase lifetime $T_2$ of $4.0\pm0.5$ ns ($40\pm4$ MHz).

The effect of a static electric field on the energy of an atom-like transition is the well-known DC Stark effect. 
It results in an energy shift of the transition which depends on the difference $\Delta\mu_\text{ind}$ between the excited and ground state dipole moments, and can be expanded as a power series in the applied electric field $F$ \cite{Boyd}.
Conventionally, Stark shift measurements focus on linear and quadratic shifts, and have been extensively used to investigate the polarizabilities of atoms and molecules \cite{Bishop1999,Buckingham2007}.
Here we consider shifts up to the fourth order in $F$, giving an expected SnV transition energy shift $\Delta E$ of
\begin{multline}
\Delta E = -\Delta\mu_\text{ind}(F)F = 
\\ = -\Delta\mu F - \frac{1}{2} \Delta\alpha F^2 - \frac{1}{3!} \Delta\beta F^3 - \frac{1}{4!} \Delta\gamma F^4
\label{eq:starkshift}
\end{multline}
The first two coefficients of this equation allow us to extract the difference between the ground and excited orbital states in permanent dipole moment $\Delta\mu$ and in polarizability $\Delta\alpha$ with the applied field, while the terms $\Delta\beta$ and $\Delta\gamma$ are related to the differences in the second and third order hyperpolarizabilities of the electronic states.
Using standard perturbation theory, the first-order Stark shift on the $i^{th}$ orbital is given by the matrix element $\bra{\psi_i}\mu\ket{\psi_i}$ where $\mu$ is the electric dipole operator, an odd function of position. Since the two ground (excited) state orbitals maintain even (odd) symmetry, these matrix elements should vanish.
The second-order correction is  given by $\sum_{j\neq i} \frac{\left|\bra{\psi_i}\mu\ket{\psi_j}\right|^2}{E_i-E_j}$. The only contributions to this sum are due to the matrix elements connecting ground and excited states, which should contribute little due to the their large energy separation.
We thus expect both the first- and second-order Stark shifts to be vanishingly small in an ideal crystal.

\begin{figure}
	\includegraphics[width=0.5\textwidth]{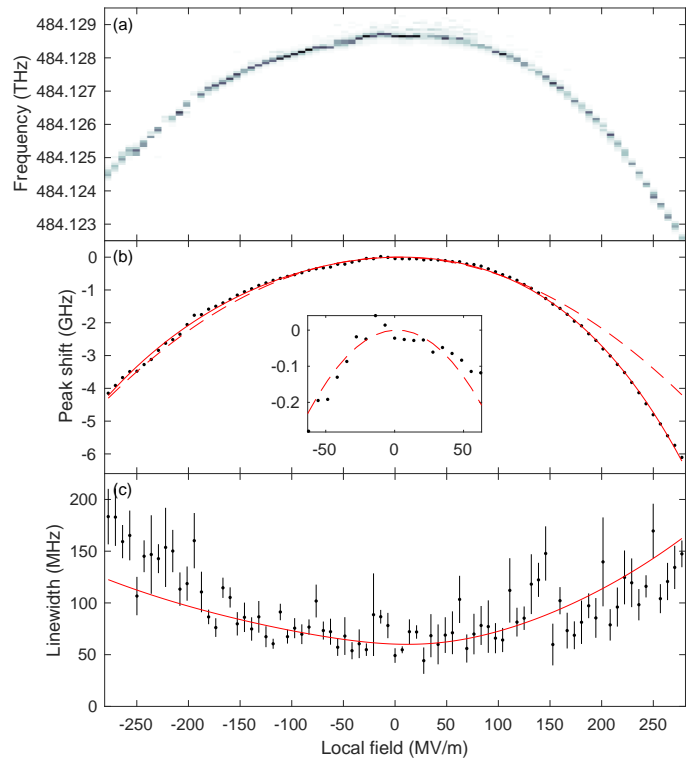}
	\caption{\label{fig:stark_shift} Stark shift of a single SnV center. (a) PLE spectra as a function of the local electric field at the defect. (b) Measured shift of the center position of the transition (black dots). The dashed and solid red lines show a fit to a second and a fourth order polynomial respectively. The inset is a close up of the shift for small applied fields, where the behaviour is quadratic. (c) Linewidth broadening caused by the induced electric dipole on the defect center.}
\end{figure}

We approximate the local electric field $F$ acting on the defect from the Lorentz local field approximation $F=F_{ext}(\epsilon+2)/3$, where $F_{ext}$ is the externally applied field, extracted from the COMSOL simulation of Figure \ref{fig:sample_geometry}(a), and $\epsilon$ the dielectric constant of diamond.
Owing to the charge stability of the defect and the high dielectric strength of diamond, we can apply high electric fields exceeding 200 MV/m and thereby detect even weak Stark effects. Figure \ref{fig:stark_shift}(a) presents the results, revealing the dependence of the SnV absorption spectrum on the electric field $F$. Such dependence clearly shows a non-linear Stark shift, a direct consequence of the absence of a permanent electric dipole in a centrosymmetric defect such as the SnV.
The black dots in Figure \ref{fig:stark_shift}(b) indicates the transition energies of the SnV emitter, determined by fitting to the PLE spectrum.
For this emitter, we extract a vanishing linear Stark shift with a slope of $6.1\times10^{-4}$ GHz/(MV/m), corresponding to a difference in dipole moment of $\Delta\mu=1.2\pm0.2\times10^{-4}$ D.
From the same fit, we can also extract a quadratic shift coefficient of $-5.1\times10^{-5}$ GHz/(MV/m)$^2$, corresponding to a polarizability difference of $\Delta\alpha=0.31\pm0.01$\AA$^3$.

\begin{figure}
	\includegraphics[width=0.33\textwidth]{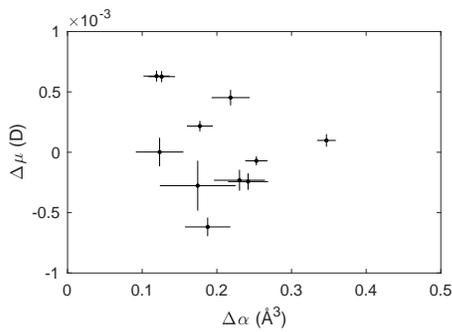}
	\caption{\label{fig:stark_parameters} Experimental values of the change in dipole moment versus the change in polarizability for 11 different SnV centers, describing respectively the first and second order Stark shift. The average polarizability is 0.23 \AA$^3$, while the average dipole moment is zero.}
\end{figure}

By studying several individual SnV centers, as summarized in Figure \ref{fig:stark_parameters}, we confirm the absence of a significant linear Stark shift in any of the observed defects, showing a distribution of $\Delta\mu$ between $-1\times10^{-3}$ D and $1\times10^{-3}$ D.
The observed polarizability difference is instead always positive with a mean of 0.23 \AA$^3$.
The measured values of $\Delta\mu$ and $\Delta\alpha$ are more than 3 orders of magnitude smaller than reported for the Stark shift of NV centers \cite{Tamarat2006,Tamarat2008}.
This result is expected from the inversion symmetry of Group IV-vacancy emitters, which leads to the absence of a permanent electric dipole in their electronic states. Additionally, the states within the ground and excited orbital manifolds share the same symmetry and are strongly split by spin-orbit coupling, which suppresses the possible electric-field induced mixing of the orbitals.
The exceptional electric field insensitivity thus arises both from the absence of a permanent electric dipole, as well as from the emitters's low polarizability.
At the same time, each emitter exhibits a Stark shift of at least 1 GHz, which corresponds to a tuning range hundreds of times larger than their natural linewidth, without any noticeable quenching of emission intensity.
Electric fields can thus be used for the spectral tuning of the transition frequency of Group IV color centers without noticeable degradation of optical properties.

The lack of a permanent dipole and low polarizability, however, is not adequate to describe the Stark shifts at high fields ($> 50$ MV/m) where higher-order corrections (hyperpolarization) become significant.
The observed dependence including these contributions is well reproduced by a fourth-order description (solid red line), showing a fit to Equation \ref{eq:starkshift}.
The third and fourth order coefficients are respectively $-5.5\pm0.3\times10^{-8}$ GHz/(MV/m)$^3$ and $-2.2\pm0.2\times10^{-10}$ GHz/(MV/m)$^4$, whose total contribution to the observed spectral trajectory is up to $25\%$.
These higher-order effects, linked to higher-order moments of the optical transition \cite{Boyd}, are typically negligible, but have been observed in systems such as molecules\cite{Latychevskaia2002} and hydrogen-like atoms, where only even-order terms are present due to their symmetry \cite{Haseyama2003}. 
This study shows that centrosymmetric defects in diamond such as the SnV enable the observation and control of hyperpolarization effects solid state systems.

To extend the analysis above, we also investigate the effect of the applied electric field on the emitter linewidth.
We fit the PLE measurements of Figure \ref{fig:stark_shift}(a) to a pseudo-Voigt profile, and extract the SnV linewidth as shown by the black dots in Figure \ref{fig:stark_shift}(c).
In these measurements the laser scan across the transition in 2.5 s, thus they will include all dephasing and spectral diffusion effects happening up to that timescale.
Without an external field, the emitter shows a narrow linewidth of $49\pm7$ MHz, which is a factor of 1.7 above the lifetime-limit of 27 MHz.
For higher fields however, a considerable broadening occurs regardless of polarity, increasing the linewidth up to 150 MHz.
We attribute this behavior to the increased sensitivity of the SnV center to charge noise in its surroundings. A higher value of a static external field $F_\text{DC}$ will induce a higher dipole on the defect, thus a same field fluctuation will shift the optical transition further.
We assume here a root mean square electric field noise at the defect location having a fixed magnitude $F_{\text{r.m.s}}\ll F_{DC}$.
By replacing $F=F_{DC}+F_{\text{r.m.s}}$ in Equation \ref{eq:starkshift}, we can see that $F_{\text{r.m.s}}$ will produce a root mean square line shift of $\sigma_G = \Delta\mu_{ind}(F_{DC}) F_{\text{r.m.s}}$.
This results in a Voigt absorption lineshape on the emitter, which can be related to the constituent Lorentzian and Gaussian width as $\Gamma_V = \frac{\Gamma_L}{2}+ \sqrt{(\frac{\Gamma_L}{2} )^2 + \Gamma_G^2}$ \cite{Whiting1968}.
We identify here $\Gamma_L$ as the homogeneous linewidth of the SnV emitter, while $\Gamma_G = 2\sqrt{2\ln2}\sigma_G$ describes the stochastic Stark shift due to $F_{\text{r.m.s}}$.
The expected SnV linewidth can then be modeled as:
\begin {equation}
\Gamma = \frac{\Gamma_{\text{L}}}{2} + \sqrt{\left(\frac{\Gamma_{\text{L}}}{2}\right)^2+8\ln2\left(F_{\text{r.m.s}} \Delta\mu_{ind}(F_{DC})\right)^2}
\label{eq:linewidth_broadening}
\end{equation}
where $\Delta\mu_{ind}(F_{DC})$ can be deduced from the Stark shift analysis of the previous section.
We fit this equation to the measured linewidth values as shown in the red line of Figure \ref{fig:stark_shift}(c), from which we extract an average field fluctuation of $F_{\text{r.m.s}} = 2.4\pm0.2$ MV/m as well as a minimum linewidth of $\Gamma_{\text{L}} = 60 \pm 4 $ MHz. Although this measured $F_{\text{r.m.s}}$ is over four orders of magnitude greater than reports for InGaAs QD devices \cite{Kuhlmann2013}, the SnV emitters still show linewidths close to the lifetime limit, highlighting again their insensitivity to the charge environment.
We note moreover that for $F_{DC}=0$ the extracted field fluctuation $F_{\text{r.m.s}}$ produces an average Stark shift much smaller than the natural linewidth of the SnV, thus the residual broadening above the lifetime limit should not be associated with electric field noise. 

\begin{figure}
	\includegraphics[width=0.5\textwidth]{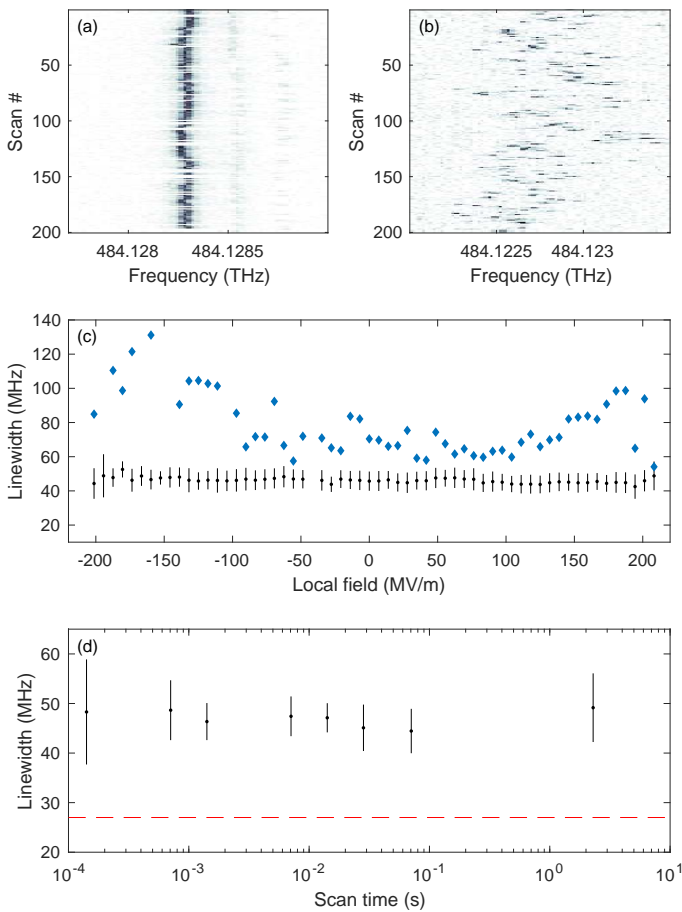}
	\caption{\label{fig:fastscan_data}  Panel (a) and (b) show the SnV transition repeatedly probed over 1.3 ms, for an electric field of 0 MV/m and 250 MV/m respectively. (c) Black dots: Lorentzian linewidth observed in an individual scan, averaged over 200 repetitions. Blue diamonds: expected Voigt linewidth by combining the Lorentzian linewidth from a single scan and the Gaussian broadening from the standard deviation of the peak positions over the 200 repetitions. (d) Single-scan linewidth measured at 0 MV/m as a function of the laser scan time across the transition (black dots). The red dashed line shows the lifetime limited value as extracted from the $g^{(2)}$ measurement of Figure 1(e).}
\end{figure}

We also investigated short-time SnV optical dynamics by fast sweeps of the PLE excitation laser at a rate of 20 GHz/s, scanning across the SnV transition in 1.3 ms.
At each bias value, the PLE scans are repeated multiple times to produce temporal series as in Figure \ref{fig:fastscan_data}(a) and (b).
These measurements reveal a narrow and stable transition at zero bias field (Figure \ref{fig:fastscan_data}(a)), while an increasingly strong spectral diffusion appears with larger electric fields (Figure \ref{fig:fastscan_data}(b)).
The mean of the PLE linewidths from individual high-speed scans as a function of the electric field is shown by the black dots in Figure \ref{fig:fastscan_data}(c) .
In contrast to the observations of Figure \ref{fig:stark_shift}(c), the linewidth measured at this timescale is consistent with a zero-bias value of 45 MHz: it does not significantly depend on the applied field.
The broadening effect can instead be reproduced from the standard deviation of the peak positions in a temporal series, as seen in the blue diamonds in Figure \ref{fig:fastscan_data}(c).
This data confirms our attribution of the SnV linewidth broadening to the electric-field dependent SnV dipole moment $\Delta\mu_{ind}(F)$, which is responsible for the stochastic Stark shift observed in Figure \ref{fig:fastscan_data}(b). 
Moreover, we can confirm that this spectral diffusion effect happens at a timescale slower than 1.3 ms.
To further investigate the residual broadening above the lifetime limit, we also measure the linewidth shown by the emitter at smaller timescales.
The results, obtained under zero bias, are shown in Figure \ref{fig:fastscan_data}(d). Here we see that there is no clear dependence on the laser scan time over the tested range, which span from 140 $\mu$s to 2.3 s. Overall, the observed linewidth is consistently a factor of 1.7 above the lifetime limit of 27 MHz.
The residual line broadening appears to originate from processes occurring above 7 kHz. Unfortunately, this timescale is not accessible here through PLE measurements due to the limited detector count rate.
However, having confirmed that the diffusion due to the stochastic Stark shifts happens at a slower rate, we can exclude charge noise as a significant contributor to broadening.
The power broadening induced by the probe laser is also negligible, so the deviation from the lifetime limited linewidth is likely due to residual interaction with the acoustic phonon bath \cite{Gorlitz2020,Jahnke2015}.

The techniques used here proved very useful to understand the effect of charge noise on SnVs.
As an additional example, we observed a linewidth narrowing effect on some SnV emitters at intermediate field values, detailed in Section 4 of the Supplementary material.
We attribute this to the suppression of spectral diffusion from the sweep-out of charge traps, which is commonly observed in other quantum emitter systems such as self-assembled quantum dots \cite{Kuhlmann2015} and silicon carbide defects \cite{Anderson2019}.  Secondly, we performed electric field-resolved spectroscopy on the 645 nm emission line that is commonly observed in Sn-implanted diamond. As detailed in section 5 of the Supplementary material, these measurements reveal a linear Stark shift, indicating a permanent dipole moment that is not consistent with the SnV's inversion symmetry.

In conclusion, we have investigated for the first time the Stark effect on a Group IV-vacancy diamond defect.
Our measurements confirm the expected first-order insensitivity to electric fields and reveal suppressed second order effects, which we quantified by the $\Delta\mu$ and $\Delta\alpha$ parameters reported in Figure \ref{fig:stark_parameters}.
Additionally, by modulating the SnV electric dipole induced by the external field, we use its linewidth to probe the local electric field noise. Using this `modulated atomic dipole' technique, we show that spectral diffusion does not significantly contribute to the SnV linewidth.
We expect this technique to be broadly useful for characterizing local electric field noise, from studies on quantum emitters \cite{Houel2012} to cold atom systems \cite{Sedlacek2018} and superconducting qubits \cite{Klimov2018}.
Our experiments also demonstrated the first Stark-shift control of a Group IV-vacancy color center emission.
Despite not offering the same tuning range as strain fields \cite{MacHielse2019}, this Stark shifting enables precise and easy-to-implement spectral tuning of Group-IV centers. It is moreover compatible with high-speed modulation, as demanded in proposals for photon-photon logic gates \cite{Heuck2020,Trivedi2020}. Combined with large-scale integration of color centers on photonic circuits \cite{Wan2019}, it opens the path to scalable precision control of quantum memories in spin-photon quantum information processing systems. 

Note: during the writing of this manuscript we become aware of similar work investigating charge fluctuations on single molecules \cite{Shkarin2020}.

L.D. acknowledge funding from the European Union’s Horizon 2020 research and innovation program under the Marie Sklodowska-Curie grant agreement No 840393.
M.T. acknowledges support through the Army Research Laboratory ENIAC Distinguished Postdoctoral Fellowship. 
K.C.C. acknowledges funding support by the National Science Foundation Graduate Research Fellowships Program (GRFP) and the Army Research Laboratory Center for Distributed Quantum Information (CDQI).
D.E. and experiments were supported in part by the STC Center for Integrated Quantum Materials (CIQM), NSF Grant No. DMR-1231319 and NSF Award No. 1839155.

\bibliography{SnVelectrodes}

\begin{thebibliography}{34}%
\makeatletter
\providecommand \@ifxundefined [1]{%
 \@ifx{#1\undefined}
}%
\providecommand \@ifnum [1]{%
 \ifnum #1\expandafter \@firstoftwo
 \else \expandafter \@secondoftwo
 \fi
}%
\providecommand \@ifx [1]{%
 \ifx #1\expandafter \@firstoftwo
 \else \expandafter \@secondoftwo
 \fi
}%
\providecommand \natexlab [1]{#1}%
\providecommand \enquote  [1]{``#1''}%
\providecommand \bibnamefont  [1]{#1}%
\providecommand \bibfnamefont [1]{#1}%
\providecommand \citenamefont [1]{#1}%
\providecommand \href@noop [0]{\@secondoftwo}%
\providecommand \href [0]{\begingroup \@sanitize@url \@href}%
\providecommand \@href[1]{\@@startlink{#1}\@@href}%
\providecommand \@@href[1]{\endgroup#1\@@endlink}%
\providecommand \@sanitize@url [0]{\catcode `\\12\catcode `\$12\catcode
  `\&12\catcode `\#12\catcode `\^12\catcode `\_12\catcode `\%12\relax}%
\providecommand \@@startlink[1]{}%
\providecommand \@@endlink[0]{}%
\providecommand \url  [0]{\begingroup\@sanitize@url \@url }%
\providecommand \@url [1]{\endgroup\@href {#1}{\urlprefix }}%
\providecommand \urlprefix  [0]{URL }%
\providecommand \Eprint [0]{\href }%
\providecommand \doibase [0]{https://doi.org/}%
\providecommand \selectlanguage [0]{\@gobble}%
\providecommand \bibinfo  [0]{\@secondoftwo}%
\providecommand \bibfield  [0]{\@secondoftwo}%
\providecommand \translation [1]{[#1]}%
\providecommand \BibitemOpen [0]{}%
\providecommand \bibitemStop [0]{}%
\providecommand \bibitemNoStop [0]{.\EOS\space}%
\providecommand \EOS [0]{\spacefactor3000\relax}%
\providecommand \BibitemShut  [1]{\csname bibitem#1\endcsname}%
\let\auto@bib@innerbib\@empty
\bibitem [{\citenamefont {Hensen}\ \emph {et~al.}(2015)\citenamefont {Hensen},
  \citenamefont {Bernien}, \citenamefont {Drea{\'{u}}}, \citenamefont
  {Reiserer}, \citenamefont {Kalb}, \citenamefont {Blok}, \citenamefont
  {Ruitenberg}, \citenamefont {Vermeulen}, \citenamefont {Schouten},
  \citenamefont {Abell{\'{a}}n}, \citenamefont {Amaya}, \citenamefont
  {Pruneri}, \citenamefont {Mitchell}, \citenamefont {Markham}, \citenamefont
  {Twitchen}, \citenamefont {Elkouss}, \citenamefont {Wehner}, \citenamefont
  {Taminiau},\ and\ \citenamefont {Hanson}}]{Hensen2015}%
  \BibitemOpen
  \bibfield  {author} {\bibinfo {author} {\bibfnamefont {B.}~\bibnamefont
  {Hensen}}, \bibinfo {author} {\bibfnamefont {H.}~\bibnamefont {Bernien}},
  \bibinfo {author} {\bibfnamefont {A.~E.}\ \bibnamefont {Drea{\'{u}}}},
  \bibinfo {author} {\bibfnamefont {A.}~\bibnamefont {Reiserer}}, \bibinfo
  {author} {\bibfnamefont {N.}~\bibnamefont {Kalb}}, \bibinfo {author}
  {\bibfnamefont {M.~S.}\ \bibnamefont {Blok}}, \bibinfo {author}
  {\bibfnamefont {J.}~\bibnamefont {Ruitenberg}}, \bibinfo {author}
  {\bibfnamefont {R.~F.}\ \bibnamefont {Vermeulen}}, \bibinfo {author}
  {\bibfnamefont {R.~N.}\ \bibnamefont {Schouten}}, \bibinfo {author}
  {\bibfnamefont {C.}~\bibnamefont {Abell{\'{a}}n}}, \bibinfo {author}
  {\bibfnamefont {W.}~\bibnamefont {Amaya}}, \bibinfo {author} {\bibfnamefont
  {V.}~\bibnamefont {Pruneri}}, \bibinfo {author} {\bibfnamefont {M.~W.}\
  \bibnamefont {Mitchell}}, \bibinfo {author} {\bibfnamefont {M.}~\bibnamefont
  {Markham}}, \bibinfo {author} {\bibfnamefont {D.~J.}\ \bibnamefont
  {Twitchen}}, \bibinfo {author} {\bibfnamefont {D.}~\bibnamefont {Elkouss}},
  \bibinfo {author} {\bibfnamefont {S.}~\bibnamefont {Wehner}}, \bibinfo
  {author} {\bibfnamefont {T.~H.}\ \bibnamefont {Taminiau}},\ and\ \bibinfo
  {author} {\bibfnamefont {R.}~\bibnamefont {Hanson}},\ }\bibfield  {title}
  {\bibinfo {title} {{Loophole-free Bell inequality violation using electron
  spins separated by 1.3 kilometres}},\ }\href
  {https://doi.org/10.1038/nature15759} {\bibfield  {journal} {\bibinfo
  {journal} {Nature}\ }\textbf {\bibinfo {volume} {526}},\ \bibinfo {pages}
  {682} (\bibinfo {year} {2015})},\ \Eprint {https://arxiv.org/abs/1508.05949}
  {arXiv:1508.05949} \BibitemShut {NoStop}%
\bibitem [{\citenamefont {Childress}\ \emph {et~al.}(2006)\citenamefont
  {Childress}, \citenamefont {{Gurudev Dutt}}, \citenamefont {Taylor},
  \citenamefont {Zibrov}, \citenamefont {Jelezko}, \citenamefont {Wrachtrup},
  \citenamefont {Hemmer},\ and\ \citenamefont {Lukin}}]{Childress2006}%
  \BibitemOpen
  \bibfield  {author} {\bibinfo {author} {\bibfnamefont {L.}~\bibnamefont
  {Childress}}, \bibinfo {author} {\bibfnamefont {M.~V.}\ \bibnamefont
  {{Gurudev Dutt}}}, \bibinfo {author} {\bibfnamefont {J.~M.}\ \bibnamefont
  {Taylor}}, \bibinfo {author} {\bibfnamefont {A.~S.}\ \bibnamefont {Zibrov}},
  \bibinfo {author} {\bibfnamefont {F.}~\bibnamefont {Jelezko}}, \bibinfo
  {author} {\bibfnamefont {J.}~\bibnamefont {Wrachtrup}}, \bibinfo {author}
  {\bibfnamefont {P.~R.}\ \bibnamefont {Hemmer}},\ and\ \bibinfo {author}
  {\bibfnamefont {M.~D.}\ \bibnamefont {Lukin}},\ }\bibfield  {title} {\bibinfo
  {title} {{Coherent dynamics of coupled electron and nuclear spin qubits in
  diamond}},\ }\href {https://doi.org/10.1126/science.1131871} {\bibfield
  {journal} {\bibinfo  {journal} {Science}\ }\textbf {\bibinfo {volume}
  {314}},\ \bibinfo {pages} {281} (\bibinfo {year} {2006})}\BibitemShut
  {NoStop}%
\bibitem [{\citenamefont {Humphreys}\ \emph {et~al.}(2018)\citenamefont
  {Humphreys}, \citenamefont {Kalb}, \citenamefont {Morits}, \citenamefont
  {Schouten}, \citenamefont {Vermeulen}, \citenamefont {Twitchen},
  \citenamefont {Markham},\ and\ \citenamefont {Hanson}}]{Humphreys2018}%
  \BibitemOpen
  \bibfield  {author} {\bibinfo {author} {\bibfnamefont {P.~C.}\ \bibnamefont
  {Humphreys}}, \bibinfo {author} {\bibfnamefont {N.}~\bibnamefont {Kalb}},
  \bibinfo {author} {\bibfnamefont {J.~P.}\ \bibnamefont {Morits}}, \bibinfo
  {author} {\bibfnamefont {R.~N.}\ \bibnamefont {Schouten}}, \bibinfo {author}
  {\bibfnamefont {R.~F.}\ \bibnamefont {Vermeulen}}, \bibinfo {author}
  {\bibfnamefont {D.~J.}\ \bibnamefont {Twitchen}}, \bibinfo {author}
  {\bibfnamefont {M.}~\bibnamefont {Markham}},\ and\ \bibinfo {author}
  {\bibfnamefont {R.}~\bibnamefont {Hanson}},\ }\bibfield  {title} {\bibinfo
  {title} {{Deterministic delivery of remote entanglement on a quantum
  network}},\ }\href {https://doi.org/10.1038/s41586-018-0200-5} {\bibfield
  {journal} {\bibinfo  {journal} {Nature}\ }\textbf {\bibinfo {volume} {558}},\
  \bibinfo {pages} {268} (\bibinfo {year} {2018})},\ \Eprint
  {https://arxiv.org/abs/1712.07567} {arXiv:1712.07567} \BibitemShut {NoStop}%
\bibitem [{\citenamefont {Gruber}\ \emph {et~al.}(1997)\citenamefont {Gruber},
  \citenamefont {Dr{\"{a}}benstedt}, \citenamefont {Tietz}, \citenamefont
  {Fleury}, \citenamefont {Wrachtrup},\ and\ \citenamefont {{Von
  Borczyskowski}}}]{Gruber1997}%
  \BibitemOpen
  \bibfield  {author} {\bibinfo {author} {\bibfnamefont {A.}~\bibnamefont
  {Gruber}}, \bibinfo {author} {\bibfnamefont {A.}~\bibnamefont
  {Dr{\"{a}}benstedt}}, \bibinfo {author} {\bibfnamefont {C.}~\bibnamefont
  {Tietz}}, \bibinfo {author} {\bibfnamefont {L.}~\bibnamefont {Fleury}},
  \bibinfo {author} {\bibfnamefont {J.}~\bibnamefont {Wrachtrup}},\ and\
  \bibinfo {author} {\bibfnamefont {C.}~\bibnamefont {{Von Borczyskowski}}},\
  }\bibfield  {title} {\bibinfo {title} {{Scanning confocal optical microscopy
  and magnetic resonance on single defect centers}},\ }\href
  {https://doi.org/10.1126/science.276.5321.2012} {\bibfield  {journal}
  {\bibinfo  {journal} {Science}\ }\textbf {\bibinfo {volume} {276}},\ \bibinfo
  {pages} {2012} (\bibinfo {year} {1997})}\BibitemShut {NoStop}%
\bibitem [{\citenamefont {Faraon}\ \emph {et~al.}(2012)\citenamefont {Faraon},
  \citenamefont {Santori}, \citenamefont {Huang}, \citenamefont {Acosta},\ and\
  \citenamefont {Beausoleil}}]{Faraon2012}%
  \BibitemOpen
  \bibfield  {author} {\bibinfo {author} {\bibfnamefont {A.}~\bibnamefont
  {Faraon}}, \bibinfo {author} {\bibfnamefont {C.}~\bibnamefont {Santori}},
  \bibinfo {author} {\bibfnamefont {Z.}~\bibnamefont {Huang}}, \bibinfo
  {author} {\bibfnamefont {V.~M.}\ \bibnamefont {Acosta}},\ and\ \bibinfo
  {author} {\bibfnamefont {R.~G.}\ \bibnamefont {Beausoleil}},\ }\bibfield
  {title} {\bibinfo {title} {{Coupling of nitrogen-vacancy centers to photonic
  crystal cavities in monocrystalline diamond}},\ }\bibfield  {journal}
  {\bibinfo  {journal} {Physical Review Letters}\ }\textbf {\bibinfo {volume}
  {109}},\ \href {https://doi.org/10.1103/PhysRevLett.109.033604}
  {10.1103/PhysRevLett.109.033604} (\bibinfo {year} {2012}),\ \Eprint
  {https://arxiv.org/abs/1202.0806} {arXiv:1202.0806} \BibitemShut {NoStop}%
\bibitem [{\citenamefont {Li}\ \emph {et~al.}(2015)\citenamefont {Li},
  \citenamefont {Schr{\"{o}}der}, \citenamefont {Chen}, \citenamefont {Walsh},
  \citenamefont {Bayn}, \citenamefont {Goldstein}, \citenamefont {Gaathon},
  \citenamefont {Trusheim}, \citenamefont {Lu}, \citenamefont {Mower},
  \citenamefont {Cotlet}, \citenamefont {Markham}, \citenamefont {Twitchen},\
  and\ \citenamefont {Englund}}]{Li2015}%
  \BibitemOpen
  \bibfield  {author} {\bibinfo {author} {\bibfnamefont {L.}~\bibnamefont
  {Li}}, \bibinfo {author} {\bibfnamefont {T.}~\bibnamefont {Schr{\"{o}}der}},
  \bibinfo {author} {\bibfnamefont {E.~H.}\ \bibnamefont {Chen}}, \bibinfo
  {author} {\bibfnamefont {M.}~\bibnamefont {Walsh}}, \bibinfo {author}
  {\bibfnamefont {I.}~\bibnamefont {Bayn}}, \bibinfo {author} {\bibfnamefont
  {J.}~\bibnamefont {Goldstein}}, \bibinfo {author} {\bibfnamefont
  {O.}~\bibnamefont {Gaathon}}, \bibinfo {author} {\bibfnamefont {M.~E.}\
  \bibnamefont {Trusheim}}, \bibinfo {author} {\bibfnamefont {M.}~\bibnamefont
  {Lu}}, \bibinfo {author} {\bibfnamefont {J.}~\bibnamefont {Mower}}, \bibinfo
  {author} {\bibfnamefont {M.}~\bibnamefont {Cotlet}}, \bibinfo {author}
  {\bibfnamefont {M.~L.}\ \bibnamefont {Markham}}, \bibinfo {author}
  {\bibfnamefont {D.~J.}\ \bibnamefont {Twitchen}},\ and\ \bibinfo {author}
  {\bibfnamefont {D.}~\bibnamefont {Englund}},\ }\bibfield  {title} {\bibinfo
  {title} {{Coherent spin control of a nanocavity-enhanced qubit in diamond}},\
  }\href {https://doi.org/10.1038/ncomms7173} {\bibfield  {journal} {\bibinfo
  {journal} {Nature Communications}\ }\textbf {\bibinfo {volume} {6}},\
  \bibinfo {pages} {6173} (\bibinfo {year} {2015})},\ \Eprint
  {https://arxiv.org/abs/1409.1602} {arXiv:1409.1602} \BibitemShut {NoStop}%
\bibitem [{\citenamefont {Evans}\ \emph {et~al.}(2016)\citenamefont {Evans},
  \citenamefont {Sipahigil}, \citenamefont {Sukachev}, \citenamefont {Zibrov},\
  and\ \citenamefont {Lukin}}]{Evans2016a}%
  \BibitemOpen
  \bibfield  {author} {\bibinfo {author} {\bibfnamefont {R.~E.}\ \bibnamefont
  {Evans}}, \bibinfo {author} {\bibfnamefont {A.}~\bibnamefont {Sipahigil}},
  \bibinfo {author} {\bibfnamefont {D.~D.}\ \bibnamefont {Sukachev}}, \bibinfo
  {author} {\bibfnamefont {A.~S.}\ \bibnamefont {Zibrov}},\ and\ \bibinfo
  {author} {\bibfnamefont {M.~D.}\ \bibnamefont {Lukin}},\ }\bibfield  {title}
  {\bibinfo {title} {{Narrow-Linewidth Homogeneous Optical Emitters in Diamond
  Nanostructures via Silicon Ion Implantation}},\ }\bibfield  {journal}
  {\bibinfo  {journal} {Physical Review Applied}\ }\textbf {\bibinfo {volume}
  {5}},\ \href {https://doi.org/10.1103/PhysRevApplied.5.044010}
  {10.1103/PhysRevApplied.5.044010} (\bibinfo {year} {2016}),\ \Eprint
  {https://arxiv.org/abs/1512.03820} {arXiv:1512.03820} \BibitemShut {NoStop}%
\bibitem [{\citenamefont {Thiering}\ and\ \citenamefont
  {Gali}(2018)}]{Thiering2018a}%
  \BibitemOpen
  \bibfield  {author} {\bibinfo {author} {\bibfnamefont {G.}~\bibnamefont
  {Thiering}}\ and\ \bibinfo {author} {\bibfnamefont {A.}~\bibnamefont
  {Gali}},\ }\bibfield  {title} {\bibinfo {title} {{Ab Initio Magneto-Optical
  Spectrum of Group-IV Vacancy Color Centers in Diamond}},\ }\href
  {https://doi.org/10.1103/PhysRevX.8.021063} {\bibfield  {journal} {\bibinfo
  {journal} {Physical Review X}\ }\textbf {\bibinfo {volume} {8}},\ \bibinfo
  {pages} {021063} (\bibinfo {year} {2018})},\ \Eprint
  {https://arxiv.org/abs/1804.07004} {arXiv:1804.07004} \BibitemShut {NoStop}%
\bibitem [{\citenamefont {Evans}\ \emph {et~al.}(2018)\citenamefont {Evans},
  \citenamefont {Bhaskar}, \citenamefont {Sukachev}, \citenamefont {Nguyen},
  \citenamefont {Sipahigil}, \citenamefont {Burek}, \citenamefont {Machielse},
  \citenamefont {Zhang}, \citenamefont {Zibrov}, \citenamefont {Bielejec},
  \citenamefont {Park}, \citenamefont {Lon{\v{c}}ar},\ and\ \citenamefont
  {Lukin}}]{Evans2018a}%
  \BibitemOpen
  \bibfield  {author} {\bibinfo {author} {\bibfnamefont {R.~E.}\ \bibnamefont
  {Evans}}, \bibinfo {author} {\bibfnamefont {M.~K.}\ \bibnamefont {Bhaskar}},
  \bibinfo {author} {\bibfnamefont {D.~D.}\ \bibnamefont {Sukachev}}, \bibinfo
  {author} {\bibfnamefont {C.~T.}\ \bibnamefont {Nguyen}}, \bibinfo {author}
  {\bibfnamefont {A.}~\bibnamefont {Sipahigil}}, \bibinfo {author}
  {\bibfnamefont {M.~J.}\ \bibnamefont {Burek}}, \bibinfo {author}
  {\bibfnamefont {B.}~\bibnamefont {Machielse}}, \bibinfo {author}
  {\bibfnamefont {G.~H.}\ \bibnamefont {Zhang}}, \bibinfo {author}
  {\bibfnamefont {A.~S.}\ \bibnamefont {Zibrov}}, \bibinfo {author}
  {\bibfnamefont {E.}~\bibnamefont {Bielejec}}, \bibinfo {author}
  {\bibfnamefont {H.}~\bibnamefont {Park}}, \bibinfo {author} {\bibfnamefont
  {M.}~\bibnamefont {Lon{\v{c}}ar}},\ and\ \bibinfo {author} {\bibfnamefont
  {M.~D.}\ \bibnamefont {Lukin}},\ }\bibfield  {title} {\bibinfo {title}
  {{Photon-mediated interactions between quantum emitters in a diamond
  nanocavity}},\ }\href {https://doi.org/10.1126/science.aau4691} {\bibfield
  {journal} {\bibinfo  {journal} {Science}\ }\textbf {\bibinfo {volume}
  {362}},\ \bibinfo {pages} {662} (\bibinfo {year} {2018})},\ \Eprint
  {https://arxiv.org/abs/1807.04265} {arXiv:1807.04265} \BibitemShut {NoStop}%
\bibitem [{\citenamefont {MacHielse}\ \emph {et~al.}(2019)\citenamefont
  {MacHielse}, \citenamefont {Bogdanovic}, \citenamefont {Meesala},
  \citenamefont {Gauthier}, \citenamefont {Burek}, \citenamefont {Joe},
  \citenamefont {Chalupnik}, \citenamefont {Sohn}, \citenamefont {Holzgrafe},
  \citenamefont {Evans}, \citenamefont {Chia}, \citenamefont {Atikian},
  \citenamefont {Bhaskar}, \citenamefont {Sukachev}, \citenamefont {Shao},
  \citenamefont {Maity}, \citenamefont {Lukin},\ and\ \citenamefont
  {Lon{\v{c}}ar}}]{MacHielse2019}%
  \BibitemOpen
  \bibfield  {author} {\bibinfo {author} {\bibfnamefont {B.}~\bibnamefont
  {MacHielse}}, \bibinfo {author} {\bibfnamefont {S.}~\bibnamefont
  {Bogdanovic}}, \bibinfo {author} {\bibfnamefont {S.}~\bibnamefont {Meesala}},
  \bibinfo {author} {\bibfnamefont {S.}~\bibnamefont {Gauthier}}, \bibinfo
  {author} {\bibfnamefont {M.~J.}\ \bibnamefont {Burek}}, \bibinfo {author}
  {\bibfnamefont {G.}~\bibnamefont {Joe}}, \bibinfo {author} {\bibfnamefont
  {M.}~\bibnamefont {Chalupnik}}, \bibinfo {author} {\bibfnamefont {Y.~I.}\
  \bibnamefont {Sohn}}, \bibinfo {author} {\bibfnamefont {J.}~\bibnamefont
  {Holzgrafe}}, \bibinfo {author} {\bibfnamefont {R.~E.}\ \bibnamefont
  {Evans}}, \bibinfo {author} {\bibfnamefont {C.}~\bibnamefont {Chia}},
  \bibinfo {author} {\bibfnamefont {H.}~\bibnamefont {Atikian}}, \bibinfo
  {author} {\bibfnamefont {M.~K.}\ \bibnamefont {Bhaskar}}, \bibinfo {author}
  {\bibfnamefont {D.~D.}\ \bibnamefont {Sukachev}}, \bibinfo {author}
  {\bibfnamefont {L.}~\bibnamefont {Shao}}, \bibinfo {author} {\bibfnamefont
  {S.}~\bibnamefont {Maity}}, \bibinfo {author} {\bibfnamefont {M.~D.}\
  \bibnamefont {Lukin}},\ and\ \bibinfo {author} {\bibfnamefont
  {M.}~\bibnamefont {Lon{\v{c}}ar}},\ }\bibfield  {title} {\bibinfo {title}
  {{Quantum Interference of Electromechanically Stabilized Emitters in
  Nanophotonic Devices}},\ }\bibfield  {journal} {\bibinfo  {journal} {Physical
  Review X}\ }\textbf {\bibinfo {volume} {9}},\ \href
  {https://doi.org/10.1103/PhysRevX.9.031022} {10.1103/PhysRevX.9.031022}
  (\bibinfo {year} {2019}),\ \Eprint {https://arxiv.org/abs/1901.09103}
  {arXiv:1901.09103} \BibitemShut {NoStop}%
\bibitem [{\citenamefont {Wan}\ \emph {et~al.}(2020)\citenamefont {Wan},
  \citenamefont {Lu}, \citenamefont {Chen}, \citenamefont {Walsh},
  \citenamefont {Trusheim}, \citenamefont {{De Santis}}, \citenamefont
  {Bersin}, \citenamefont {Harris}, \citenamefont {Mouradian}, \citenamefont
  {Christen}, \citenamefont {Bielejec},\ and\ \citenamefont
  {Englund}}]{Wan2019}%
  \BibitemOpen
  \bibfield  {author} {\bibinfo {author} {\bibfnamefont {N.~H.}\ \bibnamefont
  {Wan}}, \bibinfo {author} {\bibfnamefont {T.~J.}\ \bibnamefont {Lu}},
  \bibinfo {author} {\bibfnamefont {K.~C.}\ \bibnamefont {Chen}}, \bibinfo
  {author} {\bibfnamefont {M.~P.}\ \bibnamefont {Walsh}}, \bibinfo {author}
  {\bibfnamefont {M.~E.}\ \bibnamefont {Trusheim}}, \bibinfo {author}
  {\bibfnamefont {L.}~\bibnamefont {{De Santis}}}, \bibinfo {author}
  {\bibfnamefont {E.~A.}\ \bibnamefont {Bersin}}, \bibinfo {author}
  {\bibfnamefont {I.~B.}\ \bibnamefont {Harris}}, \bibinfo {author}
  {\bibfnamefont {S.~L.}\ \bibnamefont {Mouradian}}, \bibinfo {author}
  {\bibfnamefont {I.~R.}\ \bibnamefont {Christen}}, \bibinfo {author}
  {\bibfnamefont {E.~S.}\ \bibnamefont {Bielejec}},\ and\ \bibinfo {author}
  {\bibfnamefont {D.}~\bibnamefont {Englund}},\ }\bibfield  {title} {\bibinfo
  {title} {{Large-scale integration of artificial atoms in hybrid photonic
  circuits}},\ }\href {https://doi.org/10.1038/s41586-020-2441-3} {\bibfield
  {journal} {\bibinfo  {journal} {Nature}\ }\textbf {\bibinfo {volume} {583}},\
  \bibinfo {pages} {226} (\bibinfo {year} {2020})},\ \Eprint
  {https://arxiv.org/abs/1911.05265} {arXiv:1911.05265} \BibitemShut {NoStop}%
\bibitem [{\citenamefont {Wahl}\ \emph {et~al.}(2020)\citenamefont {Wahl},
  \citenamefont {Correia}, \citenamefont {Villarreal}, \citenamefont
  {Bourgeois}, \citenamefont {Gulka}, \citenamefont {Nesl{\'{a}}dek},
  \citenamefont {Vantomme},\ and\ \citenamefont {Pereira}}]{Wahl2020}%
  \BibitemOpen
  \bibfield  {author} {\bibinfo {author} {\bibfnamefont {U.}~\bibnamefont
  {Wahl}}, \bibinfo {author} {\bibfnamefont {J.~G.}\ \bibnamefont {Correia}},
  \bibinfo {author} {\bibfnamefont {R.}~\bibnamefont {Villarreal}}, \bibinfo
  {author} {\bibfnamefont {E.}~\bibnamefont {Bourgeois}}, \bibinfo {author}
  {\bibfnamefont {M.}~\bibnamefont {Gulka}}, \bibinfo {author} {\bibfnamefont
  {M.}~\bibnamefont {Nesl{\'{a}}dek}}, \bibinfo {author} {\bibfnamefont
  {A.}~\bibnamefont {Vantomme}},\ and\ \bibinfo {author} {\bibfnamefont
  {L.~M.}\ \bibnamefont {Pereira}},\ }\bibfield  {title} {\bibinfo {title}
  {{Direct Structural Identification and Quantification of the Split-Vacancy
  Configuration for Implanted Sn in Diamond}},\ }\href
  {https://doi.org/10.1103/PhysRevLett.125.045301} {\bibfield  {journal}
  {\bibinfo  {journal} {Physical Review Letters}\ }\textbf {\bibinfo {volume}
  {125}},\ \bibinfo {pages} {45301} (\bibinfo {year} {2020})}\BibitemShut
  {NoStop}%
\bibitem [{\citenamefont {G{\"{o}}rlitz}\ \emph {et~al.}(2020)\citenamefont
  {G{\"{o}}rlitz}, \citenamefont {Herrmann}, \citenamefont {Thiering},
  \citenamefont {Fuchs}, \citenamefont {Gandil}, \citenamefont {Iwasaki},
  \citenamefont {Taniguchi}, \citenamefont {Kieschnick}, \citenamefont
  {Meijer}, \citenamefont {Hatano}, \citenamefont {Gali},\ and\ \citenamefont
  {Becher}}]{Gorlitz2020}%
  \BibitemOpen
  \bibfield  {author} {\bibinfo {author} {\bibfnamefont {J.}~\bibnamefont
  {G{\"{o}}rlitz}}, \bibinfo {author} {\bibfnamefont {D.}~\bibnamefont
  {Herrmann}}, \bibinfo {author} {\bibfnamefont {G.}~\bibnamefont {Thiering}},
  \bibinfo {author} {\bibfnamefont {P.}~\bibnamefont {Fuchs}}, \bibinfo
  {author} {\bibfnamefont {M.}~\bibnamefont {Gandil}}, \bibinfo {author}
  {\bibfnamefont {T.}~\bibnamefont {Iwasaki}}, \bibinfo {author} {\bibfnamefont
  {T.}~\bibnamefont {Taniguchi}}, \bibinfo {author} {\bibfnamefont
  {M.}~\bibnamefont {Kieschnick}}, \bibinfo {author} {\bibfnamefont
  {J.}~\bibnamefont {Meijer}}, \bibinfo {author} {\bibfnamefont
  {M.}~\bibnamefont {Hatano}}, \bibinfo {author} {\bibfnamefont
  {A.}~\bibnamefont {Gali}},\ and\ \bibinfo {author} {\bibfnamefont
  {C.}~\bibnamefont {Becher}},\ }\bibfield  {title} {\bibinfo {title}
  {{Spectroscopic investigations of negatively charged tin-vacancy centres in
  diamond}},\ }\href {https://doi.org/10.1088/1367-2630/ab6631} {\bibfield
  {journal} {\bibinfo  {journal} {New Journal of Physics}\ }\textbf {\bibinfo
  {volume} {22}},\ \bibinfo {pages} {13048} (\bibinfo {year} {2020})},\ \Eprint
  {https://arxiv.org/abs/1909.09435} {arXiv:1909.09435} \BibitemShut {NoStop}%
\bibitem [{\citenamefont {Rugar}\ \emph {et~al.}(2020)\citenamefont {Rugar},
  \citenamefont {Lu}, \citenamefont {Dory}, \citenamefont {Sun}, \citenamefont
  {McQuade}, \citenamefont {Shen}, \citenamefont {Melosh},\ and\ \citenamefont
  {Vu{\v{c}}kovi{\'{c}}}}]{Rugar2020}%
  \BibitemOpen
  \bibfield  {author} {\bibinfo {author} {\bibfnamefont {A.~E.}\ \bibnamefont
  {Rugar}}, \bibinfo {author} {\bibfnamefont {H.}~\bibnamefont {Lu}}, \bibinfo
  {author} {\bibfnamefont {C.}~\bibnamefont {Dory}}, \bibinfo {author}
  {\bibfnamefont {S.}~\bibnamefont {Sun}}, \bibinfo {author} {\bibfnamefont
  {P.~J.}\ \bibnamefont {McQuade}}, \bibinfo {author} {\bibfnamefont {Z.~X.}\
  \bibnamefont {Shen}}, \bibinfo {author} {\bibfnamefont {N.~A.}\ \bibnamefont
  {Melosh}},\ and\ \bibinfo {author} {\bibfnamefont {J.}~\bibnamefont
  {Vu{\v{c}}kovi{\'{c}}}},\ }\bibfield  {title} {\bibinfo {title} {{Generation
  of Tin-Vacancy Centers in Diamond via Shallow Ion Implantation and Subsequent
  Diamond Overgrowth}},\ }\href {https://doi.org/10.1021/acs.nanolett.9b04495}
  {\bibfield  {journal} {\bibinfo  {journal} {Nano Letters}\ }\textbf {\bibinfo
  {volume} {20}},\ \bibinfo {pages} {1614} (\bibinfo {year} {2020})},\ \Eprint
  {https://arxiv.org/abs/1910.14165} {arXiv:1910.14165} \BibitemShut {NoStop}%
\bibitem [{\citenamefont {Iwasaki}\ \emph {et~al.}(2017)\citenamefont
  {Iwasaki}, \citenamefont {Miyamoto}, \citenamefont {Taniguchi}, \citenamefont
  {Siyushev}, \citenamefont {Metsch}, \citenamefont {Jelezko},\ and\
  \citenamefont {Hatano}}]{Iwasaki2017}%
  \BibitemOpen
  \bibfield  {author} {\bibinfo {author} {\bibfnamefont {T.}~\bibnamefont
  {Iwasaki}}, \bibinfo {author} {\bibfnamefont {Y.}~\bibnamefont {Miyamoto}},
  \bibinfo {author} {\bibfnamefont {T.}~\bibnamefont {Taniguchi}}, \bibinfo
  {author} {\bibfnamefont {P.}~\bibnamefont {Siyushev}}, \bibinfo {author}
  {\bibfnamefont {M.~H.}\ \bibnamefont {Metsch}}, \bibinfo {author}
  {\bibfnamefont {F.}~\bibnamefont {Jelezko}},\ and\ \bibinfo {author}
  {\bibfnamefont {M.}~\bibnamefont {Hatano}},\ }\bibfield  {title} {\bibinfo
  {title} {{Tin-Vacancy Quantum Emitters in Diamond}},\ }\bibfield  {journal}
  {\bibinfo  {journal} {Physical Review Letters}\ }\textbf {\bibinfo {volume}
  {119}},\ \href {https://doi.org/10.1103/PhysRevLett.119.253601}
  {10.1103/PhysRevLett.119.253601} (\bibinfo {year} {2017}),\ \Eprint
  {https://arxiv.org/abs/1708.03576} {arXiv:1708.03576} \BibitemShut {NoStop}%
\bibitem [{\citenamefont {Trusheim}\ \emph {et~al.}(2020)\citenamefont
  {Trusheim}, \citenamefont {Pingault}, \citenamefont {Wan}, \citenamefont
  {G{\"{u}}ndoǧan}, \citenamefont {{De Santis}}, \citenamefont {Debroux},
  \citenamefont {Gangloff}, \citenamefont {Purser}, \citenamefont {Chen},
  \citenamefont {Walsh}, \citenamefont {Rose}, \citenamefont {Becker},
  \citenamefont {Lienhard}, \citenamefont {Bersin}, \citenamefont
  {Paradeisanos}, \citenamefont {Wang}, \citenamefont {Lyzwa}, \citenamefont
  {Montblanch}, \citenamefont {Malladi}, \citenamefont {Bakhru}, \citenamefont
  {Ferrari}, \citenamefont {Walmsley}, \citenamefont {Atat{\"{u}}re},\ and\
  \citenamefont {Englund}}]{Trusheim2020}%
  \BibitemOpen
  \bibfield  {author} {\bibinfo {author} {\bibfnamefont {M.~E.}\ \bibnamefont
  {Trusheim}}, \bibinfo {author} {\bibfnamefont {B.}~\bibnamefont {Pingault}},
  \bibinfo {author} {\bibfnamefont {N.~H.}\ \bibnamefont {Wan}}, \bibinfo
  {author} {\bibfnamefont {M.}~\bibnamefont {G{\"{u}}ndoǧan}}, \bibinfo
  {author} {\bibfnamefont {L.}~\bibnamefont {{De Santis}}}, \bibinfo {author}
  {\bibfnamefont {R.}~\bibnamefont {Debroux}}, \bibinfo {author} {\bibfnamefont
  {D.}~\bibnamefont {Gangloff}}, \bibinfo {author} {\bibfnamefont
  {C.}~\bibnamefont {Purser}}, \bibinfo {author} {\bibfnamefont {K.~C.}\
  \bibnamefont {Chen}}, \bibinfo {author} {\bibfnamefont {M.}~\bibnamefont
  {Walsh}}, \bibinfo {author} {\bibfnamefont {J.~J.}\ \bibnamefont {Rose}},
  \bibinfo {author} {\bibfnamefont {J.~N.}\ \bibnamefont {Becker}}, \bibinfo
  {author} {\bibfnamefont {B.}~\bibnamefont {Lienhard}}, \bibinfo {author}
  {\bibfnamefont {E.}~\bibnamefont {Bersin}}, \bibinfo {author} {\bibfnamefont
  {I.}~\bibnamefont {Paradeisanos}}, \bibinfo {author} {\bibfnamefont
  {G.}~\bibnamefont {Wang}}, \bibinfo {author} {\bibfnamefont {D.}~\bibnamefont
  {Lyzwa}}, \bibinfo {author} {\bibfnamefont {A.~R.}\ \bibnamefont
  {Montblanch}}, \bibinfo {author} {\bibfnamefont {G.}~\bibnamefont {Malladi}},
  \bibinfo {author} {\bibfnamefont {H.}~\bibnamefont {Bakhru}}, \bibinfo
  {author} {\bibfnamefont {A.~C.}\ \bibnamefont {Ferrari}}, \bibinfo {author}
  {\bibfnamefont {I.~A.}\ \bibnamefont {Walmsley}}, \bibinfo {author}
  {\bibfnamefont {M.}~\bibnamefont {Atat{\"{u}}re}},\ and\ \bibinfo {author}
  {\bibfnamefont {D.}~\bibnamefont {Englund}},\ }\bibfield  {title} {\bibinfo
  {title} {{Transform-Limited Photons from a Coherent Tin-Vacancy Spin in
  Diamond}},\ }\href {https://doi.org/10.1103/PhysRevLett.124.023602}
  {\bibfield  {journal} {\bibinfo  {journal} {Physical Review Letters}\
  }\textbf {\bibinfo {volume} {124}},\ \bibinfo {pages} {23602} (\bibinfo
  {year} {2020})},\ \Eprint {https://arxiv.org/abs/1811.07777}
  {arXiv:1811.07777} \BibitemShut {NoStop}%
\bibitem [{\citenamefont {Bertolotti}(2015)}]{Boyd}%
  \BibitemOpen
  \bibfield  {author} {\bibinfo {author} {\bibfnamefont {M.}~\bibnamefont
  {Bertolotti}},\ }\href {https://doi.org/10.1201/b18201-9} {\emph {\bibinfo
  {title} {Masers and Lasers}}}\ (\bibinfo {year} {2015})\ pp.\ \bibinfo
  {pages} {108--161}\BibitemShut {NoStop}%
\bibitem [{\citenamefont {Bishop}(1999)}]{Bishop1999}%
  \BibitemOpen
  \bibfield  {author} {\bibinfo {author} {\bibfnamefont {D.~M.}\ \bibnamefont
  {Bishop}},\ }\bibfield  {title} {\bibinfo {title} {{Polarizability and
  hyperpolarizability of atoms and ions}},\ }\href
  {https://doi.org/10.1016/S1380-7323(99)80007-7} {\bibfield  {journal}
  {\bibinfo  {journal} {Theoretical and Computational Chemistry}\ }\textbf
  {\bibinfo {volume} {6}},\ \bibinfo {pages} {129} (\bibinfo {year}
  {1999})}\BibitemShut {NoStop}%
\bibitem [{\citenamefont {Buckingham}(2007)}]{Buckingham2007}%
  \BibitemOpen
  \bibfield  {author} {\bibinfo {author} {\bibfnamefont {A.~D.}\ \bibnamefont
  {Buckingham}},\ }\bibfield  {title} {\bibinfo {title} {{Permanent and Induced
  Molecular Moments and Long-Range Intermolecular Forces}},\ }in\ \href
  {https://doi.org/10.1002/9780470143582.ch2} {\emph {\bibinfo {booktitle}
  {Advances in Chemical Physics, 1967, 12, 107 - 142}}}\ (\bibinfo  {publisher}
  {John Wiley anf Sons, Ltd},\ \bibinfo {year} {2007})\ pp.\ \bibinfo {pages}
  {107--142}\BibitemShut {NoStop}%
\bibitem [{\citenamefont {Tamarat}\ \emph {et~al.}(2006)\citenamefont
  {Tamarat}, \citenamefont {Gaebel}, \citenamefont {Rabeau}, \citenamefont
  {Khan}, \citenamefont {Greentree}, \citenamefont {Wilson}, \citenamefont
  {Hollenberg}, \citenamefont {Prawer}, \citenamefont {Hemmer}, \citenamefont
  {Jelezko},\ and\ \citenamefont {Wrachtrup}}]{Tamarat2006}%
  \BibitemOpen
  \bibfield  {author} {\bibinfo {author} {\bibfnamefont {P.}~\bibnamefont
  {Tamarat}}, \bibinfo {author} {\bibfnamefont {T.}~\bibnamefont {Gaebel}},
  \bibinfo {author} {\bibfnamefont {J.~R.}\ \bibnamefont {Rabeau}}, \bibinfo
  {author} {\bibfnamefont {M.}~\bibnamefont {Khan}}, \bibinfo {author}
  {\bibfnamefont {A.~D.}\ \bibnamefont {Greentree}}, \bibinfo {author}
  {\bibfnamefont {H.}~\bibnamefont {Wilson}}, \bibinfo {author} {\bibfnamefont
  {L.~C.}\ \bibnamefont {Hollenberg}}, \bibinfo {author} {\bibfnamefont
  {S.}~\bibnamefont {Prawer}}, \bibinfo {author} {\bibfnamefont
  {P.}~\bibnamefont {Hemmer}}, \bibinfo {author} {\bibfnamefont
  {F.}~\bibnamefont {Jelezko}},\ and\ \bibinfo {author} {\bibfnamefont
  {J.}~\bibnamefont {Wrachtrup}},\ }\bibfield  {title} {\bibinfo {title}
  {{Stark shift control of single optical centers in diamond}},\ }\bibfield
  {journal} {\bibinfo  {journal} {Physical Review Letters}\ }\textbf {\bibinfo
  {volume} {97}},\ \href {https://doi.org/10.1103/PhysRevLett.97.083002}
  {10.1103/PhysRevLett.97.083002} (\bibinfo {year} {2006})\BibitemShut
  {NoStop}%
\bibitem [{\citenamefont {Tamarat}\ \emph {et~al.}(2008)\citenamefont
  {Tamarat}, \citenamefont {Manson}, \citenamefont {Harrison}, \citenamefont
  {McMurtrie}, \citenamefont {Nizovtsev}, \citenamefont {Santori},
  \citenamefont {Beausoleil}, \citenamefont {Neumann}, \citenamefont {Gaebel},
  \citenamefont {Jelezko}, \citenamefont {Hemmer},\ and\ \citenamefont
  {Wrachtrup}}]{Tamarat2008}%
  \BibitemOpen
  \bibfield  {author} {\bibinfo {author} {\bibfnamefont {P.}~\bibnamefont
  {Tamarat}}, \bibinfo {author} {\bibfnamefont {N.~B.}\ \bibnamefont {Manson}},
  \bibinfo {author} {\bibfnamefont {J.~P.}\ \bibnamefont {Harrison}}, \bibinfo
  {author} {\bibfnamefont {R.~L.}\ \bibnamefont {McMurtrie}}, \bibinfo {author}
  {\bibfnamefont {A.}~\bibnamefont {Nizovtsev}}, \bibinfo {author}
  {\bibfnamefont {C.}~\bibnamefont {Santori}}, \bibinfo {author} {\bibfnamefont
  {R.~G.}\ \bibnamefont {Beausoleil}}, \bibinfo {author} {\bibfnamefont
  {P.}~\bibnamefont {Neumann}}, \bibinfo {author} {\bibfnamefont
  {T.}~\bibnamefont {Gaebel}}, \bibinfo {author} {\bibfnamefont
  {F.}~\bibnamefont {Jelezko}}, \bibinfo {author} {\bibfnamefont
  {P.}~\bibnamefont {Hemmer}},\ and\ \bibinfo {author} {\bibfnamefont
  {J.}~\bibnamefont {Wrachtrup}},\ }\bibfield  {title} {\bibinfo {title}
  {{Spin-flip and spin-conserving optical transitions of the nitrogen-vacancy
  centre in diamond}},\ }\bibfield  {journal} {\bibinfo  {journal} {New Journal
  of Physics}\ }\textbf {\bibinfo {volume} {10}},\ \href
  {https://doi.org/10.1088/1367-2630/10/4/045004}
  {10.1088/1367-2630/10/4/045004} (\bibinfo {year} {2008})\BibitemShut
  {NoStop}%
\bibitem [{\citenamefont {Latychevskaia}\ \emph {et~al.}(2002)\citenamefont
  {Latychevskaia}, \citenamefont {Renn},\ and\ \citenamefont
  {Wild}}]{Latychevskaia2002}%
  \BibitemOpen
  \bibfield  {author} {\bibinfo {author} {\bibfnamefont {T.~Y.}\ \bibnamefont
  {Latychevskaia}}, \bibinfo {author} {\bibfnamefont {A.}~\bibnamefont
  {Renn}},\ and\ \bibinfo {author} {\bibfnamefont {U.~P.}\ \bibnamefont
  {Wild}},\ }\bibfield  {title} {\bibinfo {title} {{Higher-order Stark effect
  on single-molecules}},\ }\href
  {https://doi.org/10.1016/S0301-0104(02)00621-3} {\bibfield  {journal}
  {\bibinfo  {journal} {Chemical Physics}\ }\textbf {\bibinfo {volume} {282}},\
  \bibinfo {pages} {109} (\bibinfo {year} {2002})}\BibitemShut {NoStop}%
\bibitem [{\citenamefont {Haseyama}\ \emph {et~al.}(2003)\citenamefont
  {Haseyama}, \citenamefont {Kominato}, \citenamefont {Shibata}, \citenamefont
  {Yamada}, \citenamefont {Saida}, \citenamefont {Nakura}, \citenamefont
  {Kishimoto}, \citenamefont {Tada}, \citenamefont {Ogawa}, \citenamefont
  {Funahashi}, \citenamefont {Yamamoto},\ and\ \citenamefont
  {Matsuki}}]{Haseyama2003}%
  \BibitemOpen
  \bibfield  {author} {\bibinfo {author} {\bibfnamefont {T.}~\bibnamefont
  {Haseyama}}, \bibinfo {author} {\bibfnamefont {K.}~\bibnamefont {Kominato}},
  \bibinfo {author} {\bibfnamefont {M.}~\bibnamefont {Shibata}}, \bibinfo
  {author} {\bibfnamefont {S.}~\bibnamefont {Yamada}}, \bibinfo {author}
  {\bibfnamefont {T.}~\bibnamefont {Saida}}, \bibinfo {author} {\bibfnamefont
  {T.}~\bibnamefont {Nakura}}, \bibinfo {author} {\bibfnamefont
  {Y.}~\bibnamefont {Kishimoto}}, \bibinfo {author} {\bibfnamefont
  {M.}~\bibnamefont {Tada}}, \bibinfo {author} {\bibfnamefont {I.}~\bibnamefont
  {Ogawa}}, \bibinfo {author} {\bibfnamefont {H.}~\bibnamefont {Funahashi}},
  \bibinfo {author} {\bibfnamefont {K.}~\bibnamefont {Yamamoto}},\ and\
  \bibinfo {author} {\bibfnamefont {S.}~\bibnamefont {Matsuki}},\ }\bibfield
  {title} {\bibinfo {title} {{Second- and fourth-order Stark shifts and their
  principal-quantum-number dependence in high Rydberg states of 85Rb}},\ }\href
  {https://doi.org/10.1016/j.physleta.2003.09.011} {\bibfield  {journal}
  {\bibinfo  {journal} {Physics Letters, Section A}\ }\textbf {\bibinfo
  {volume} {317}},\ \bibinfo {pages} {450} (\bibinfo {year}
  {2003})}\BibitemShut {NoStop}%
\bibitem [{\citenamefont {Whiting}(1968)}]{Whiting1968}%
  \BibitemOpen
  \bibfield  {author} {\bibinfo {author} {\bibfnamefont {E.~E.}\ \bibnamefont
  {Whiting}},\ }\bibfield  {title} {\bibinfo {title} {{An empirical
  approximation to the Voigt profile}},\ }\href
  {https://doi.org/10.1016/0022-4073(68)90081-2} {\bibfield  {journal}
  {\bibinfo  {journal} {Journal of Quantitative Spectroscopy and Radiative
  Transfer}\ }\textbf {\bibinfo {volume} {8}},\ \bibinfo {pages} {1379}
  (\bibinfo {year} {1968})}\BibitemShut {NoStop}%
\bibitem [{\citenamefont {Kuhlmann}\ \emph {et~al.}(2013)\citenamefont
  {Kuhlmann}, \citenamefont {Houel}, \citenamefont {Ludwig}, \citenamefont
  {Greuter}, \citenamefont {Reuter}, \citenamefont {Wieck}, \citenamefont
  {Poggio},\ and\ \citenamefont {Warburton}}]{Kuhlmann2013}%
  \BibitemOpen
  \bibfield  {author} {\bibinfo {author} {\bibfnamefont {A.~V.}\ \bibnamefont
  {Kuhlmann}}, \bibinfo {author} {\bibfnamefont {J.}~\bibnamefont {Houel}},
  \bibinfo {author} {\bibfnamefont {A.}~\bibnamefont {Ludwig}}, \bibinfo
  {author} {\bibfnamefont {L.}~\bibnamefont {Greuter}}, \bibinfo {author}
  {\bibfnamefont {D.}~\bibnamefont {Reuter}}, \bibinfo {author} {\bibfnamefont
  {A.~D.}\ \bibnamefont {Wieck}}, \bibinfo {author} {\bibfnamefont
  {M.}~\bibnamefont {Poggio}},\ and\ \bibinfo {author} {\bibfnamefont {R.~J.}\
  \bibnamefont {Warburton}},\ }\bibfield  {title} {\bibinfo {title} {{Charge
  noise and spin noise in a semiconductor quantum device}},\ }\href
  {https://doi.org/10.1038/nphys2688} {\bibfield  {journal} {\bibinfo
  {journal} {Nature Physics}\ }\textbf {\bibinfo {volume} {9}},\ \bibinfo
  {pages} {570} (\bibinfo {year} {2013})},\ \Eprint
  {https://arxiv.org/abs/1301.6381} {arXiv:1301.6381} \BibitemShut {NoStop}%
\bibitem [{\citenamefont {Jahnke}\ \emph {et~al.}(2015)\citenamefont {Jahnke},
  \citenamefont {Sipahigil}, \citenamefont {Binder}, \citenamefont {Doherty},
  \citenamefont {Metsch}, \citenamefont {Rogers}, \citenamefont {Manson},
  \citenamefont {Lukin},\ and\ \citenamefont {Jelezko}}]{Jahnke2015}%
  \BibitemOpen
  \bibfield  {author} {\bibinfo {author} {\bibfnamefont {K.~D.}\ \bibnamefont
  {Jahnke}}, \bibinfo {author} {\bibfnamefont {A.}~\bibnamefont {Sipahigil}},
  \bibinfo {author} {\bibfnamefont {J.~M.}\ \bibnamefont {Binder}}, \bibinfo
  {author} {\bibfnamefont {M.~W.}\ \bibnamefont {Doherty}}, \bibinfo {author}
  {\bibfnamefont {M.}~\bibnamefont {Metsch}}, \bibinfo {author} {\bibfnamefont
  {L.~J.}\ \bibnamefont {Rogers}}, \bibinfo {author} {\bibfnamefont {N.~B.}\
  \bibnamefont {Manson}}, \bibinfo {author} {\bibfnamefont {M.~D.}\
  \bibnamefont {Lukin}},\ and\ \bibinfo {author} {\bibfnamefont
  {F.}~\bibnamefont {Jelezko}},\ }\bibfield  {title} {\bibinfo {title}
  {{Electron-phonon processes of the silicon-vacancy centre in diamond}},\
  }\bibfield  {journal} {\bibinfo  {journal} {New Journal of Physics}\ }\textbf
  {\bibinfo {volume} {17}},\ \href
  {https://doi.org/10.1088/1367-2630/17/4/043011}
  {10.1088/1367-2630/17/4/043011} (\bibinfo {year} {2015}),\ \Eprint
  {https://arxiv.org/abs/1411.2871} {arXiv:1411.2871} \BibitemShut {NoStop}%
\bibitem [{\citenamefont {Kuhlmann}\ \emph {et~al.}(2015)\citenamefont
  {Kuhlmann}, \citenamefont {Prechtel}, \citenamefont {Houel}, \citenamefont
  {Ludwig}, \citenamefont {Reuter}, \citenamefont {Wieck},\ and\ \citenamefont
  {Warburton}}]{Kuhlmann2015}%
  \BibitemOpen
  \bibfield  {author} {\bibinfo {author} {\bibfnamefont {A.~V.}\ \bibnamefont
  {Kuhlmann}}, \bibinfo {author} {\bibfnamefont {J.~H.}\ \bibnamefont
  {Prechtel}}, \bibinfo {author} {\bibfnamefont {J.}~\bibnamefont {Houel}},
  \bibinfo {author} {\bibfnamefont {A.}~\bibnamefont {Ludwig}}, \bibinfo
  {author} {\bibfnamefont {D.}~\bibnamefont {Reuter}}, \bibinfo {author}
  {\bibfnamefont {A.~D.}\ \bibnamefont {Wieck}},\ and\ \bibinfo {author}
  {\bibfnamefont {R.~J.}\ \bibnamefont {Warburton}},\ }\bibfield  {title}
  {\bibinfo {title} {{Transform-limited single photons from a single quantum
  dot}},\ }\href {https://doi.org/10.1038/ncomms9204} {\bibfield  {journal}
  {\bibinfo  {journal} {Nature Communications}\ }\textbf {\bibinfo {volume}
  {6}},\ \bibinfo {pages} {8204} (\bibinfo {year} {2015})},\ \Eprint
  {https://arxiv.org/abs/1307.7109} {arXiv:1307.7109} \BibitemShut {NoStop}%
\bibitem [{\citenamefont {Anderson}\ \emph {et~al.}(2019)\citenamefont
  {Anderson}, \citenamefont {Bourassa}, \citenamefont {Miao}, \citenamefont
  {Wolfowicz}, \citenamefont {Mintun}, \citenamefont {Crook}, \citenamefont
  {Abe}, \citenamefont {{Ul Hassan}}, \citenamefont {Son}, \citenamefont
  {Ohshima},\ and\ \citenamefont {Awschalom}}]{Anderson2019}%
  \BibitemOpen
  \bibfield  {author} {\bibinfo {author} {\bibfnamefont {C.~P.}\ \bibnamefont
  {Anderson}}, \bibinfo {author} {\bibfnamefont {A.}~\bibnamefont {Bourassa}},
  \bibinfo {author} {\bibfnamefont {K.~C.}\ \bibnamefont {Miao}}, \bibinfo
  {author} {\bibfnamefont {G.}~\bibnamefont {Wolfowicz}}, \bibinfo {author}
  {\bibfnamefont {P.~J.}\ \bibnamefont {Mintun}}, \bibinfo {author}
  {\bibfnamefont {A.~L.}\ \bibnamefont {Crook}}, \bibinfo {author}
  {\bibfnamefont {H.}~\bibnamefont {Abe}}, \bibinfo {author} {\bibfnamefont
  {J.}~\bibnamefont {{Ul Hassan}}}, \bibinfo {author} {\bibfnamefont {N.~T.}\
  \bibnamefont {Son}}, \bibinfo {author} {\bibfnamefont {T.}~\bibnamefont
  {Ohshima}},\ and\ \bibinfo {author} {\bibfnamefont {D.~D.}\ \bibnamefont
  {Awschalom}},\ }\bibfield  {title} {\bibinfo {title} {{Electrical and optical
  control of single spins integrated in scalable semiconductor devices}},\
  }\href {https://doi.org/10.1126/science.aax9406} {\bibfield  {journal}
  {\bibinfo  {journal} {Science}\ }\textbf {\bibinfo {volume} {366}},\ \bibinfo
  {pages} {1225} (\bibinfo {year} {2019})},\ \Eprint
  {https://arxiv.org/abs/1906.08328} {arXiv:1906.08328} \BibitemShut {NoStop}%
\bibitem [{\citenamefont {Houel}\ \emph {et~al.}(2012)\citenamefont {Houel},
  \citenamefont {Kuhlmann}, \citenamefont {Greuter}, \citenamefont {Xue},
  \citenamefont {Poggio}, \citenamefont {Warburton}, \citenamefont {Gerardot},
  \citenamefont {Dalgarno}, \citenamefont {Badolato}, \citenamefont {Petroff},
  \citenamefont {Ludwig}, \citenamefont {Reuter},\ and\ \citenamefont
  {Wieck}}]{Houel2012}%
  \BibitemOpen
  \bibfield  {author} {\bibinfo {author} {\bibfnamefont {J.}~\bibnamefont
  {Houel}}, \bibinfo {author} {\bibfnamefont {A.~V.}\ \bibnamefont {Kuhlmann}},
  \bibinfo {author} {\bibfnamefont {L.}~\bibnamefont {Greuter}}, \bibinfo
  {author} {\bibfnamefont {F.}~\bibnamefont {Xue}}, \bibinfo {author}
  {\bibfnamefont {M.}~\bibnamefont {Poggio}}, \bibinfo {author} {\bibfnamefont
  {R.~J.}\ \bibnamefont {Warburton}}, \bibinfo {author} {\bibfnamefont {B.~D.}\
  \bibnamefont {Gerardot}}, \bibinfo {author} {\bibfnamefont {P.~A.}\
  \bibnamefont {Dalgarno}}, \bibinfo {author} {\bibfnamefont {A.}~\bibnamefont
  {Badolato}}, \bibinfo {author} {\bibfnamefont {P.~M.}\ \bibnamefont
  {Petroff}}, \bibinfo {author} {\bibfnamefont {A.}~\bibnamefont {Ludwig}},
  \bibinfo {author} {\bibfnamefont {D.}~\bibnamefont {Reuter}},\ and\ \bibinfo
  {author} {\bibfnamefont {A.~D.}\ \bibnamefont {Wieck}},\ }\bibfield  {title}
  {\bibinfo {title} {{Probing single-charge fluctuations at a GaAs/AlAs
  interface using laser spectroscopy on a nearby InGaAs quantum dot}},\ }\href
  {https://doi.org/10.1103/PhysRevLett.108.107401} {\bibfield  {journal}
  {\bibinfo  {journal} {Physical Review Letters}\ }\textbf {\bibinfo {volume}
  {108}},\ \bibinfo {pages} {1} (\bibinfo {year} {2012})},\ \Eprint
  {https://arxiv.org/abs/1110.2714v1} {arXiv:1110.2714v1} \BibitemShut
  {NoStop}%
\bibitem [{\citenamefont {Sedlacek}\ \emph {et~al.}(2018)\citenamefont
  {Sedlacek}, \citenamefont {Greene}, \citenamefont {Stuart}, \citenamefont
  {McConnell}, \citenamefont {Bruzewicz}, \citenamefont {Sage},\ and\
  \citenamefont {Chiaverini}}]{Sedlacek2018}%
  \BibitemOpen
  \bibfield  {author} {\bibinfo {author} {\bibfnamefont {J.~A.}\ \bibnamefont
  {Sedlacek}}, \bibinfo {author} {\bibfnamefont {A.}~\bibnamefont {Greene}},
  \bibinfo {author} {\bibfnamefont {J.}~\bibnamefont {Stuart}}, \bibinfo
  {author} {\bibfnamefont {R.}~\bibnamefont {McConnell}}, \bibinfo {author}
  {\bibfnamefont {C.~D.}\ \bibnamefont {Bruzewicz}}, \bibinfo {author}
  {\bibfnamefont {J.~M.}\ \bibnamefont {Sage}},\ and\ \bibinfo {author}
  {\bibfnamefont {J.}~\bibnamefont {Chiaverini}},\ }\bibfield  {title}
  {\bibinfo {title} {{Distance scaling of electric-field noise in a
  surface-electrode ion trap}},\ }\bibfield  {journal} {\bibinfo  {journal}
  {Physical Review A}\ }\textbf {\bibinfo {volume} {97}},\ \href
  {https://doi.org/10.1103/PhysRevA.97.020302} {10.1103/PhysRevA.97.020302}
  (\bibinfo {year} {2018}),\ \Eprint {https://arxiv.org/abs/1712.00188}
  {arXiv:1712.00188} \BibitemShut {NoStop}%
\bibitem [{\citenamefont {Klimov}\ \emph {et~al.}(2018)\citenamefont {Klimov},
  \citenamefont {Kelly}, \citenamefont {Chen}, \citenamefont {Neeley},
  \citenamefont {Megrant}, \citenamefont {Burkett}, \citenamefont {Barends},
  \citenamefont {Arya}, \citenamefont {Chiaro}, \citenamefont {Chen},
  \citenamefont {Dunsworth}, \citenamefont {Fowler}, \citenamefont {Foxen},
  \citenamefont {Gidney}, \citenamefont {Giustina}, \citenamefont {Graff},
  \citenamefont {Huang}, \citenamefont {Jeffrey}, \citenamefont {Lucero},
  \citenamefont {Mutus}, \citenamefont {Naaman}, \citenamefont {Neill},
  \citenamefont {Quintana}, \citenamefont {Roushan}, \citenamefont {Sank},
  \citenamefont {Vainsencher}, \citenamefont {Wenner}, \citenamefont {White},
  \citenamefont {Boixo}, \citenamefont {Babbush}, \citenamefont {Smelyanskiy},
  \citenamefont {Neven},\ and\ \citenamefont {Martinis}}]{Klimov2018}%
  \BibitemOpen
  \bibfield  {author} {\bibinfo {author} {\bibfnamefont {P.~V.}\ \bibnamefont
  {Klimov}}, \bibinfo {author} {\bibfnamefont {J.}~\bibnamefont {Kelly}},
  \bibinfo {author} {\bibfnamefont {Z.}~\bibnamefont {Chen}}, \bibinfo {author}
  {\bibfnamefont {M.}~\bibnamefont {Neeley}}, \bibinfo {author} {\bibfnamefont
  {A.}~\bibnamefont {Megrant}}, \bibinfo {author} {\bibfnamefont
  {B.}~\bibnamefont {Burkett}}, \bibinfo {author} {\bibfnamefont
  {R.}~\bibnamefont {Barends}}, \bibinfo {author} {\bibfnamefont
  {K.}~\bibnamefont {Arya}}, \bibinfo {author} {\bibfnamefont {B.}~\bibnamefont
  {Chiaro}}, \bibinfo {author} {\bibfnamefont {Y.}~\bibnamefont {Chen}},
  \bibinfo {author} {\bibfnamefont {A.}~\bibnamefont {Dunsworth}}, \bibinfo
  {author} {\bibfnamefont {A.}~\bibnamefont {Fowler}}, \bibinfo {author}
  {\bibfnamefont {B.}~\bibnamefont {Foxen}}, \bibinfo {author} {\bibfnamefont
  {C.}~\bibnamefont {Gidney}}, \bibinfo {author} {\bibfnamefont
  {M.}~\bibnamefont {Giustina}}, \bibinfo {author} {\bibfnamefont
  {R.}~\bibnamefont {Graff}}, \bibinfo {author} {\bibfnamefont
  {T.}~\bibnamefont {Huang}}, \bibinfo {author} {\bibfnamefont
  {E.}~\bibnamefont {Jeffrey}}, \bibinfo {author} {\bibfnamefont
  {E.}~\bibnamefont {Lucero}}, \bibinfo {author} {\bibfnamefont {J.~Y.}\
  \bibnamefont {Mutus}}, \bibinfo {author} {\bibfnamefont {O.}~\bibnamefont
  {Naaman}}, \bibinfo {author} {\bibfnamefont {C.}~\bibnamefont {Neill}},
  \bibinfo {author} {\bibfnamefont {C.}~\bibnamefont {Quintana}}, \bibinfo
  {author} {\bibfnamefont {P.}~\bibnamefont {Roushan}}, \bibinfo {author}
  {\bibfnamefont {D.}~\bibnamefont {Sank}}, \bibinfo {author} {\bibfnamefont
  {A.}~\bibnamefont {Vainsencher}}, \bibinfo {author} {\bibfnamefont
  {J.}~\bibnamefont {Wenner}}, \bibinfo {author} {\bibfnamefont {T.~C.}\
  \bibnamefont {White}}, \bibinfo {author} {\bibfnamefont {S.}~\bibnamefont
  {Boixo}}, \bibinfo {author} {\bibfnamefont {R.}~\bibnamefont {Babbush}},
  \bibinfo {author} {\bibfnamefont {V.~N.}\ \bibnamefont {Smelyanskiy}},
  \bibinfo {author} {\bibfnamefont {H.}~\bibnamefont {Neven}},\ and\ \bibinfo
  {author} {\bibfnamefont {J.~M.}\ \bibnamefont {Martinis}},\ }\bibfield
  {title} {\bibinfo {title} {{Fluctuations of Energy-Relaxation Times in
  Superconducting Qubits}},\ }\bibfield  {journal} {\bibinfo  {journal}
  {Physical Review Letters}\ }\textbf {\bibinfo {volume} {121}},\ \href
  {https://doi.org/10.1103/PhysRevLett.121.090502}
  {10.1103/PhysRevLett.121.090502} (\bibinfo {year} {2018})\BibitemShut
  {NoStop}%
\bibitem [{\citenamefont {Heuck}\ \emph {et~al.}(2020)\citenamefont {Heuck},
  \citenamefont {Jacobs},\ and\ \citenamefont {Englund}}]{Heuck2020}%
  \BibitemOpen
  \bibfield  {author} {\bibinfo {author} {\bibfnamefont {M.}~\bibnamefont
  {Heuck}}, \bibinfo {author} {\bibfnamefont {K.}~\bibnamefont {Jacobs}},\ and\
  \bibinfo {author} {\bibfnamefont {D.~R.}\ \bibnamefont {Englund}},\
  }\bibfield  {title} {\bibinfo {title} {{Photon-photon interactions in
  dynamically coupled cavities}},\ }\href
  {https://doi.org/10.1103/PhysRevA.101.042322} {\bibfield  {journal} {\bibinfo
   {journal} {Physical Review A}\ }\textbf {\bibinfo {volume} {101}},\ \bibinfo
  {pages} {42322} (\bibinfo {year} {2020})},\ \Eprint
  {https://arxiv.org/abs/1905.02134} {arXiv:1905.02134} \BibitemShut {NoStop}%
\bibitem [{\citenamefont {Trivedi}\ \emph {et~al.}(2020)\citenamefont
  {Trivedi}, \citenamefont {White}, \citenamefont {Fan},\ and\ \citenamefont
  {Vu{\v{c}}kovi{\'{c}}}}]{Trivedi2020}%
  \BibitemOpen
  \bibfield  {author} {\bibinfo {author} {\bibfnamefont {R.}~\bibnamefont
  {Trivedi}}, \bibinfo {author} {\bibfnamefont {A.}~\bibnamefont {White}},
  \bibinfo {author} {\bibfnamefont {S.}~\bibnamefont {Fan}},\ and\ \bibinfo
  {author} {\bibfnamefont {J.}~\bibnamefont {Vu{\v{c}}kovi{\'{c}}}},\
  }\bibfield  {title} {\bibinfo {title} {{Analytic and geometric properties of
  scattering from periodically modulated quantum-optical systems}},\ }\bibfield
   {journal} {\bibinfo  {journal} {Physical Review A}\ }\textbf {\bibinfo
  {volume} {102}},\ \href {https://doi.org/10.1103/PhysRevA.102.033707}
  {10.1103/PhysRevA.102.033707} (\bibinfo {year} {2020}),\ \Eprint
  {https://arxiv.org/abs/2003.10673} {arXiv:2003.10673} \BibitemShut {NoStop}%
\bibitem [{\citenamefont {Shkarin}\ \emph {et~al.}(2020)\citenamefont
  {Shkarin}, \citenamefont {Rattenbacher}, \citenamefont {Renger},
  \citenamefont {H{\"{o}}nl}, \citenamefont {Utikal}, \citenamefont {Seidler},
  \citenamefont {G{\"{o}}tzinger},\ and\ \citenamefont
  {Sandoghdar}}]{Shkarin2020}%
  \BibitemOpen
  \bibfield  {author} {\bibinfo {author} {\bibfnamefont {A.}~\bibnamefont
  {Shkarin}}, \bibinfo {author} {\bibfnamefont {D.}~\bibnamefont
  {Rattenbacher}}, \bibinfo {author} {\bibfnamefont {J.}~\bibnamefont
  {Renger}}, \bibinfo {author} {\bibfnamefont {S.}~\bibnamefont {H{\"{o}}nl}},
  \bibinfo {author} {\bibfnamefont {T.}~\bibnamefont {Utikal}}, \bibinfo
  {author} {\bibfnamefont {P.}~\bibnamefont {Seidler}}, \bibinfo {author}
  {\bibfnamefont {S.}~\bibnamefont {G{\"{o}}tzinger}},\ and\ \bibinfo {author}
  {\bibfnamefont {V.}~\bibnamefont {Sandoghdar}},\ }\bibfield  {title}
  {\bibinfo {title} {{Nanoscopic charge fluctuations in a gallium phosphide
  waveguide measured by single molecules}},\ }\href
  {http://arxiv.org/abs/2012.02991} {\  (\bibinfo {year} {2020})},\ \Eprint
  {https://arxiv.org/abs/2012.02991} {arXiv:2012.02991} \BibitemShut {NoStop}%
\end{thebibliography}%

\end{document}